\documentclass[a4paper,nobibnotes,nofootinbib]{revtex4}

\usepackage{amsmath,amssymb}
\usepackage{epsfig}
\usepackage{graphicx}
\usepackage{xspace}

\newcommand{\be}{\begin{equation}}
\newcommand{\ee}{\end{equation}}
\newcommand{\bL}{\begin{Large}}
\newcommand{\eL}{\end{Large}}
\newcommand{\ba}{\begin{eqnarray}}
\newcommand{\ea}{\end{eqnarray}}
\newcommand{\bc}{\begin{center}}
\newcommand{\ec}{\end{center}}
\newcommand{\bfig}{\begin{figure}}
\newcommand{\efig}{\end{figure}}

\newcommand{\la}{\label}

\newcommand{\rr}[4]{#1, {\it #2 \/}{\bf #3} #4}

\newcommand{\chic}{\chi_{c_0}}
\newcommand{\chib}{\chi_{b_0}}
\newcommand{\chitwo}{\chi_{2}}

\begin{document}

\title {
Diffractive $\chi$ Production at the Tevatron and 
the LHC}
\author{M. Rangel}\email{rangel@cbpf.br}
\affiliation{LAFEX, Centro Brasileiro de Pesquisas Fis{\'i}cas, Rio de Janeiro, 
Brazil}

\author{C. Royon}\email{royon@hep.saclay.cea.fr}
\affiliation{Service de physique des particules, CEA/Saclay,
  91191 Gif-sur-Yvette cedex, France}

\author{G. Alves}\email{gilvan@cbpf.br}
\affiliation{LAFEX, Centro Brasileiro de Pesquisas Fis{\'i}cas, Rio de Janeiro, 
Brazil }

\author{J. Barreto}\email{barreto@fnal.gov}
\affiliation{Instituto de F\' {i}sica da Universidade Federal do Rio de 
Janeiro, Brazil}

\author{and R. Peschanski}\email{pesch@spht.saclay.cea.fr}
\affiliation{Service de physique th{\'e}orique, CEA/Saclay,
  91191 Gif-sur-Yvette cedex, France\footnote{%
URA 2306, unit{\'e} de recherche associ{\'e}e au CNRS.}}

\begin{abstract}
We present predictions for the diffractive production of $\chi$ mesons in the 
central rapidity region usually covered by collider detectors. The predicted
cross sections are based on the Bialas-Landshoff formalism for both exclusive 
and 
inclusive production and makes use of the 
DPEMC Monte-Carlo simulation adapted with kinematics appropriate for small-mass 
diffractive production. We compare generator-level results with a CDF 
measurement for exclusive $\chi$ production, and study background and other 
scenarios 
including the contribution 
of inclusive $\chi$ production.  The results agree with  the Tevatron data 
and are extrapolated, highlighting the exclusive $\chic$ production 
at LHC energies. A possible new measurement at the Tevatron using the D\O\ 
forward detectors is investigated, taking advantage of  the dominance of 
exclusive production for high enough diffractive mass fraction.
\end{abstract}

\maketitle

\section{\textbf{Introduction}}
\la{I}
Exclusive and inclusive central diffractive production of heavy states have 
been 
studied 
previously in the double
Pomeron exchange formalism (DPE) 
\cite{bialas1,others,bialas2,khoze,us,cox,sci,us1}, 
and experimental results have been presented 
\cite{kn:cdf}, attracting theoretical attention. Indeed, massive dijets have 
been copiously produced in inclusive diffractive production at the Tevatron and 
there is presently an active search for exclusive  diffractive production of 
heavy states. One motivation for this search is that the Higgs boson could be 
produced in such a mode, allowing for a good mass determination for this 
elusive 
particle if such exclusive events are identified \cite{al00}. However the 
estimated 
cross sections and signal-over-noise ratios are still a matter of debate.

One way to address this problem is looking for a similar production mechanism 
with 
lighter particles like the $\chi$ mesons \cite{chic}. This would give rise to high enough 
cross sections 
to check the dynamical 
mechanisms. Indeed, exclusive production of $\chi_{c}$ has been reported by the 
CDF collaboration \cite{kn:michele} with an upper limit for the cross section 
of 
\begin{equation} \label{eq:uppercdf}
\sigma_{exc}(p\bar{p} \rightarrow p+J/\psi
 + \gamma+\bar{p}) < 49 \pm 18 (stat)
\pm 39 (sys)\  pb\ .
\end{equation}

Our goal in the present paper is to analyse both exclusive and inclusive 
diffractive production of $\chi_{c_0}$ and $\chi_{b_0}$ in the context of the 
Tevatron, making useful predictions for the LHC. 
A possible new measurement using the D\O\ forward detectors is also 
investigated. 
As a general framework, 
 we use an 
extension to low mass states of the Bialas-Landshoff (BL)  model 
\cite{bialas1} for 
both exclusive and inclusive production. More precisely, we consider  the  
inclusive model as applied in Ref.\cite{us} and the extension to purely 
exclusive 
processes \cite{us1},
using the corresponding color-singlet ($J_z$ = 0) subprocess cross sections of 
the original Bialas-Landshoff formalism for diffractive 
Higgs 
production \cite{bialas1}, applied also to heavy quark pairs in 
\cite{bialas2}. In fact, in our approach, both inclusive and exclusive diagrams 
come from
the same non-perturbative approach, originated in \cite{bialas1}.
The model for exclusive production starts with the 
same soft Pomeron exchange diagrams  (for ordinary dijet production  the $gg\to 
gg$ diagrams are also included \cite{bzdak})  and corrects the 
result by  non-perturbative  rapidity-gap survival factors 
\cite{sp,Kupco:2004fw}. For inclusive production in the same framework, no such 
factor is applied, but the normalisation is fixed by fitting the CDF dijet data 
\cite{kn:cdf}. 
The 
energy 
dependence 
is related to the rise of ordinary hadronic cross-sections through
features of the soft Pomeron \cite{pom}. We think that the non-perturbative 
Bialas-Landshoff approach is  well suited for the production of relatively 
low-mass diffractive $\chi$ states\footnote{At this point, only $\chi_{c0}$ and 
$\chi_{b0}$
mesons can be evaluated. Indeed, due to non-zero $p_{\perp}$ of the 
initial state
(pp or p$\bar{p}$),
there will be contributions of $\chi_{c_2}$ and $\chi_{b_2}$ in the exclusive
amplitude. This will be discussed in a further paper.}.

The full kinematics is evaluated for low-mass exclusive production and 
implemented
in the DPEMC Monte Carlo generator \cite{kn:dpemc}, originally designed for 
heavy 
mass states. We use the generator level distributions to estimate the 
measurements 
at detector level, and to study the systematic uncertainties due the 
lack of a full detector simulation.

The paper is divided in the following way: in section \ref{II}, we recall 
briefly 
the  Bialas-Landshoff (BL) formalism for both exclusive and inclusive 
diffractive production and give the theoretical cross section formulae. They 
include the full diffractive kinematics for a low mass state, which is derived 
in section \ref{III} and implemented in the DPEMC Monte Carlo. In section 
\ref{IV}, 
we show the predictions for the exclusive and inclusive diffractive $\chi$ 
production 
cross sections at the Tevatron and LHC.  In the subsequent section  \ref{V} we 
compare 
our predictions with the CDF measurement and investigate the possibility of a 
new 
measurement at the 
D\O\  detector in section \ref{VI}. Section  \ref{VII} is devoted to the 
predictions for 
exclusive $\chi_{c0}$ production at the LHC and in section \ref{VIII} we 
present 
our conclusions 
and an outlook on this interesting mode of production.

\section{\textbf{The Bialas-Landshoff formalism}}
\la{II}

One generally considers two types of DPE topologies for the production of a 
heavy 
state: exclusive DPE 
\cite{bialas1,others,bialas2,khoze,us1}, where the central  object is produced
alone, separated from the outgoing hadrons by rapidity gaps: 
\be h h \rightarrow h + \text{heavy object} + h\ , \label{exc}\ee 
\noindent and inclusive DPE \cite{others,us,cox,sci,khoze}, where the 
colliding Pomerons are resolved (very much like ordinary hadrons), accompanying
the central object with Pomeron ``remnants'' (X,Y):
\be h h \rightarrow h + X + \text{heavy object} + Y + h\ . \label{inc}\ee 
\noindent in both cases $h$ represents the colliding hadrons.

In general, exclusive production is considered most promising, since a 
better signal-to-background ratio is expected.
Indeed, 
if the events are exclusive, \emph{i.e.}, no other particles are produced 
in addition to the heavy object and the outgoing hadrons, the measurement
of the scattered hadrons in near-beam detectors provides information on the 
mass 
of the heavy object \cite{al00}, and the dynamics of the hard 
process. However, the approximations made for 
heavy mass states lead to inconsistent results in the case of lighter objects, 
so
in our case we have to develop the full kinematics for all mass states.

In order to evaluate both inclusive and exclusive diffractive production, we 
use an extension \cite{us1} to the purely exclusive 
processes 
 of the original Bialas-Landshoff formalism for diffractive Higgs 
production \cite{bialas1}, also applied to heavy quark pairs
\cite{bialas2}. In this extension, both inclusive and exclusive diagrams come 
from
the same approach and are based on  soft Pomeron exchange diagrams  (for 
ordinary dijet production  the $gg\to 
gg$ diagrams are also included \cite{bzdak}). In the exclusive case one has 
to   correct the 
result by  non-perturbative  rapidity-gap survival factors 
\cite{sp,Kupco:2004fw}, while in the inclusive case the normalisation is fixed 
\cite{us} by reference with dijet production \cite{kn:cdf}. The values
of the survival probabilities are the same as for the diffractive Higgs
production in Ref. \cite{us1}, namely 0.1 and 0.03 for respectively the 
Tevatron and the LHC.
The energy dependence of the cross section 
is related to the rise of ordinary hadronic cross sections through
features of the soft Pomeron \cite{pom}.

With respect to previous works, the formulae of the Bialas-Landshoff 
cross sections have to include now the full kinematics for diffractive $\chi$ 
meson states and thus valid for small masses in general. This kinematics 
is explained in the next section.  For the 
exclusive 
$\chi$ meson 
production in hadron-hadron collisions, 
one has
\begin{equation} \label{eq:exc}
d\sigma_{\chi}^{exc}(s) =  C_{\chi} \left ( \frac{s}{{M_{\chi}^2}}
 \right )
^{2\epsilon}
\delta \left [ \frac{{M_{\chi}}^2}{s} -  \frac{M^2_{diff}}{s}
\right ]  \ \prod_{i=1,2} \left \{ d^2v_i \ \frac{d\xi_i}{1-\xi_i} \  
\xi_i^{2\alpha'v_i^2}\ 
\exp(-2\lambda_{\chi}v_i^2) \right \} \ ,
\end{equation}
where $C_{\chi}$ is the normalisation constant (cf. \cite{bialas1} for the 
Higgs boson), $M_{\chi}$ is the $\chi$ meson mass and  $\lambda_{\chi}$ the 
slope at the Pomeron vertex. 
$\alpha'$ and $\epsilon$ 
are the standard soft Pomeron parameters \cite{pom}. The gluon-gluon
$\chi_C$ coupling satisfying the $J_z=0$ rule
is  taken as in Ref. \cite{chic}.
The parameter $C_{\chi}$ 
contains a non perturbative part due to the gluon
coupling $G$ to the proton. We kept the original value in the model of Bialas
Landshoff which means $G^2/4 \pi \sim 1$. This is of course an order of magnitude,
and it leads to an uncertainty on the exclusive production cross section.

Most importantly, $M^2_{diff}$ is the expression of the diffractive mass 
produced in the central region
in terms of the  kinematic variables. For low-mass states, as the $\chi's$, it 
sensibly differs from the expression for heavy diffractive states, as will 
be derived in the next section and in Appendix \ref{App}. This is one major 
modification one has to introduce w.r.t. the original formalism.

For inclusive production, the cross sections are given by 
\begin{eqnarray} \label{eq:inc}
d\sigma_{\chi}^{inc}(s) & = & C^{inc}_{\chi} \left ( \frac{x_1^g 
x_2^g}{{M_{\chi}^2}}
 \right )
^{2\epsilon}
\delta \left ( \xi_1 \xi_2 - \frac{M_{\chi}^2}{x_1^g x_2^gs} \right ) \\
\nonumber
& &
\ \ \ \ \ \ \ \ \prod_{i=1,2} \left \{ G_p(x_i^g,\mu)\ d^2v_i dx_i^g \ 
\frac{d\xi_i}{1-\xi_i}
\ \xi_i^{2\alpha^`v_i^2}
\ exp(-2\lambda_{\chi}v_i^2) \right \} \ , 
\end{eqnarray}
where one makes use of the parton structure functions $ G_p(x_i^g,\mu)$ in the 
Pomeron, for a given scale $\mu$. Note that in Equation \eqref{eq:inc},
the normalisation $C^{inc}_{\chi}$ is fixed by normalising the prediction on the
measurement of the dijet diffractive cross section by the CDF collaboration
\cite{us}.

In both equations,  the variables $v_{i}$ and $\xi_{i}$  
denote the transverse
momenta and fractional momentum losses of the outgoing hadrons. In the second 
equation, 
$x^g_{i}$ denote the fractional momentum carried by the gluons in the 
Pomeron. Note that in the inclusive case, the usual kinematics is used for 
$M_{diff}^2\equiv \frac{M_{\chi}^2}{x_1^g x_2^g}.$
\section{\textbf{Full kinematics for exclusive production}}
\la{III}
Let us first recall the method used to generate exclusive events in the DPEMC
Monte Carlo generator \cite{kn:dpemc}. The first step is to randomly generate 
$t_1$, 
$t_2$ and $\xi_1$, following an exponential distribution for $|t_i|$, where 
$|t_i|$ is the 4-momentum
transferred  and $\xi_i$ is the momentum loss for the hadron $i$. Exclusive
events have the property that the full energy available in the center-of-mass 
is
used to produce the diffractive object, or in other words there is no
Pomeron remnant. The diffractive mass can be expressed as:

\begin{eqnarray}
M^2_{diff} \approx s \xi_1 \xi_2.
\label{mass}
\end{eqnarray}

The value of $\xi_2$ is thus
imposed by this relationship. The produced events in the generator are then
weighted as usual by the cross section.

The approximation \eqref{mass} is no longer true for low mass states
such as $\chi$ mesons,
and we had to modify the method to generate events in this case.
We derived the diffractive mass from full 
4-momentum conservation (see Appendix A).

Using the full kinematics, Equation \eqref{mass} is replaced by

\begin{equation} \label{eq:diffmass}
M^2_{diff} = s \times \left ( 1 + 
\frac{(1-\xi_1)(1-\xi_2)}{2cos\theta_1
cos\theta_2}(1-\Omega) - \left ( \frac{1-\xi_1}{cos\theta_1} +
\frac{1-\xi_2}{cos\theta_2} \right ) \right )\ ,
\end{equation}
where $\Omega = -cos{\theta_1} cos{\theta_2} + sin{\theta_1} 
sin{\theta_2}(cos{\varphi_1}cos{\varphi_2}+sin{\varphi_1}
sin{\varphi_2})$, $\theta$ is the scattering angle and $\varphi$ 
the polar angle.
It is important to notice that this formula  depends not only on  
$\xi_1$ and $\xi_2$ but also on the angles of the hadrons $\theta_1,\varphi_1$ 
and
$\theta_2,\varphi_2.$ $t$ and $\theta$ are related by the following formula:

\begin{equation} \label{eq:theta}
\sin^2 \theta_{1,2} \sim \theta_{1,2}^2 = \frac{|t_{1,2}|}{(1-\xi_{1,2})(s/4)}.
\end{equation}
We use the following method to generate low mass exclusive events within the
framework of DPMEC \cite{kn:dpemc}. We start by generating $\theta_1$,
$\theta_2$ (following an exponential distribution) 
and $\xi_1$ randomly, which gives $t_1$ by the equation \eqref{eq:theta}.
$\xi_2$ is then computed inverting equation \eqref{eq:diffmass}, giving also
$t_2$. The events are then weighted according to
the cross section. The new steps are thus to use the variables $\theta$ and 
$\xi$, 
avoiding the cumbersome solution of equation \eqref{eq:diffmass} in terms of 
$\xi$
and $t$.

\section{\textbf{Exclusive and inclusive $\chic$, $\chib$ production
cross sections}}
\begin{figure}[ht]\begin{center}
\epsfig{file=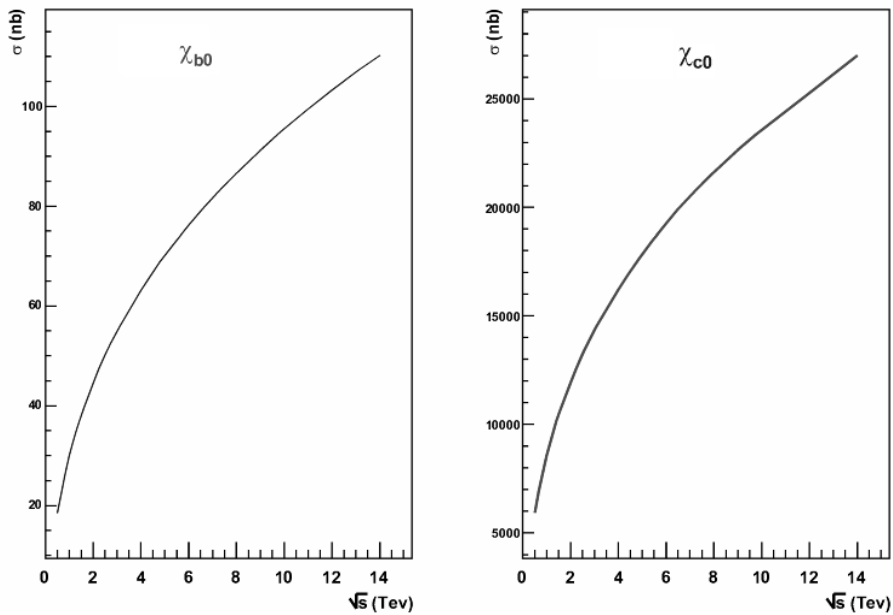, width=0.7\textwidth}
\end{center}
\caption{\small{Exclusive Production cross section of 
$\chi_{b0}$ (left) and $\chi_{c0}$ mesons (right). }
\label{fig:cross}}
\end{figure}

\la{IV}
Table \eqref{tab:cross} presents our results for the cross section predictions 
at
the Tevatron and
the LHC. The energy dependence for the exclusive production cross section can 
be 
seen in
the Fig. \ref{fig:cross} for the 
$\chi_{b_0}$ (left) and $\chi_{c_0}$ (right) mesons.
The gap survival probability (the probability of the gaps not
to be populated) $S_{gap}^2$ is taken to be 
0.1 at the Tevatron  and 
0.03 at the LHC (cf. A.B. Kaidalov et al in \cite{sp}).
\begin{table}[hbtp]
\begin{center}
\begin{tabular}{|c|c|c|} 
  \hline
  $\sigma (nb) $& Tevatron $\sqrt{s}=1.96$ TeV & LHC $\sqrt{s}=14$ TeV  \\
   
  \hline
  $\sigma_{exc}$($\chi_{c_0}$) & 1.17$\times10^3$ & 0.804$\times10^3$ \\
  $\sigma_{exc}$($\chi_{b_0}$) & 4.4 & 3.29 \\
  $\sigma_{inc}$($\chi_{c_0}$) & 1.8 $\times10^4$ & 4.8 $\times10^4$ \\
  $\sigma_{inc}$($\chi_{b_0}$) & 20 & 1.8 $\times10^2$ \\
  \hline
\end{tabular}
\caption{\small{Cross sections (in nb) for exclusive and inclusive production 
at the Tevatron and the LHC.}\label{tab:cross}}
\end{center}
\end{table}

Note that our estimates for exclusive $\chi_{c_0}$
production at the Tevatron are higher than other 
recent predictions \cite{chic}. The differences arise mainly because of the 
distinct Pomeron fluxes in the models, which means
that the BL exclusive cross section has a smoother dependence with 
the center of mass energy ($\sqrt{s}$) (see Fig. \ref{fig:cross}). 
We should also note that our model does not contain  Sudakov factors
contrary to Ref. \cite{khoze, chic}. The effect of the Sudakov suppression is
however supposed to be small at small masses, and plays a more
important role at higher masses.

Possible sources of background for exclusive production include:
\begin{itemize}
\item  cosmic events which can fake an exclusiveness;
\item  exclusive and inclusive $\chitwo$ production;
\item  quasi-exclusive $\chi$ production, which are defined by a high mass
fraction F$_M$ (e.g. F$_M$ $\geq$ 0.75-0.95),
\end{itemize}
where the mass fraction F$_M$ is the ratio between  $M_{\chi}$ and the total 
diffractive
mass $M_{diff}$. We will study this last contamination to the exclusive signal 
in the
following.

\section{\textbf{Comparison with the CDF limits on $\chi_{c_0}$ Production at 
the Tevatron}}
\la{V}

As shown in table \ref{tab:cross}, exclusive $\chi_{c_0}$ production
at the Tevatron has a noticeable cross section, with the potential
to be detected in the traditional decay $\chi_{c_0} \rightarrow J/\psi 
(\rightarrow \mu^+ +  \mu^-)+ \gamma$. The CDF Collaboration
has presented preliminary results \cite{kn:michele} for exclusive 
$J/\psi + \gamma$ production using the rapidity gap selection of
diffractive events in Run II ($\sqrt{s} = $1.96 TeV). The cuts used by CDF 
are the following: $p_T (\mu^{\pm}) \geq 1.5$ (GeV),
$|\eta (\gamma)| \leq 3.5$ and $|\eta (\mu^{\pm})| \leq 0.6$.
The CDF collaboration 
provided an upper limit on $\chi_{c_0}$
production cross section at the Tevatron, uncorrected for residual
backgrounds such as cosmics \footnote{This cross section has been corrected for
the $J/\Psi$ branching ratio into muons.}:

\begin{equation} \label{eq:upper}
\sigma_{exc}(p\bar{p} \rightarrow p+J/\psi
 + \gamma+\bar{p}) < 49 \pm 18 (stat)
\pm 39 (sys)\  pb.
\end{equation}

If we apply the CDF cuts at generator level, we predict the following cross
section 

\begin{equation} \label{eq:chiccdf}
\sigma_{exc}(p\bar{p} \rightarrow p+\chi_{c0} (\rightarrow J/\psi 
\gamma) + \bar{p}) = 61  pb.
\end{equation}

In order to make a more realistic comparison, we need also to consider the 
non-exclusive 
background which can enter directly
in the experimental cross section determination. In particular, the 
contamination 
due to quasi-exclusive events need to be considered properly.
In this class of events the
QCD radiation, or in other words the energy loss in the Pomeron remnant and in
soft QCD radiation, is small. CDF remove most of the inclusive
background using a cut on the mass fraction, F$_M > 0.85$. For exclusive 
events, 
F$_M$
is expected to be close to one, since the full available energy is used to
produce the $\chi$ meson, and smaller than one for inclusive events. 
However, this cut is of course applied at the detector
level by the CDF collaboration, whereas we can only apply it at the generator
level. Due to the fact that we are missing the smearing between detector and
generator levels, we choose to investigate the effect on the cross section due 
to 
various
mass-fraction cuts, as displayed in Table \ref{tab:quasitev}.

\begin{figure}[h]\begin{center}
\epsfig{file=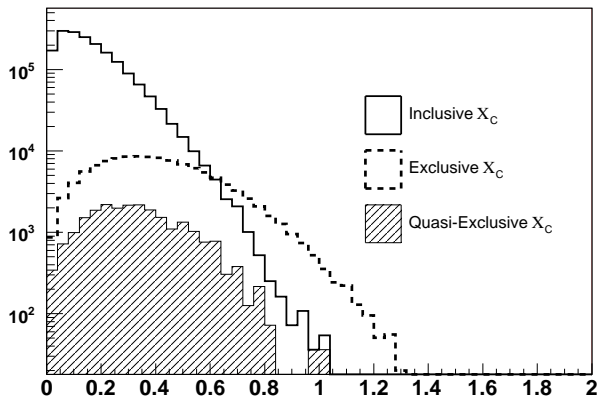, width=0.5\textwidth}
\end{center}
\caption{\small{Transverse momentum of $\chi_{c0}$ for exclusive production
(solid line), inclusive production (dashed line) and for quasi-
inclusive events (filled) at $\sqrt{s}$
 = 1.96 TeV (the normalization used a luminosity of 100 $nb^{-1}$).
}
\label{fig:pT}}
\end{figure}

We also compared the transverse momentum (p$_T$) distributions of the $\chic$ 
for 
inclusive,
exclusive and quasi-exclusive production, the latest defined as F$_M \geq$ 0.75
(Fig. \ref{fig:pT}). As expected,
the exclusive p$_T$ distribution reaches higher values, so this variable 
could be used to enhance the exclusive production
signal.

The quasi-exclusive contamination to the exclusive signal also suffers from the
unknown gluon density in the Pomeron at high $\beta$ in particular. This 
uncertainty, estimated to be about 50\% \cite{laurent}, can be taken into 
account 
by multiplying the gluon density
in the Pomeron, measured at HERA, by a factor $(1-\beta)^{\nu}$  where $\nu$
varies between -1.0 and 1.0 \cite{olda}. If $\nu$ is negative, we enhance the 
gluon density
at high $\beta$ by definition. The QCD fits to the HERA data lead to the
following value of the $\nu$ parameter: $\nu = 0.0 \pm 0.6$.

In summary, the sources of contamination to the exclusive signal can derive 
from
the smearing in the mass fraction, which is a pure experimental effect, or the
uncertainties in the gluon density in the Pomeron, which is a theoretical 
limitation.

Table II gives the quasi-exclusive cross section at the Tevatron after applying
the CDF cuts, and different cuts on the mass fraction and values of the
$\nu$ parameer.  We thus find 
that the signal seen by the CDF collaboration could be
explained by a combination of a higher gluon density at high $\beta$ and some 
smearing effects due
to the reconstruction of the mass fraction. Table III gives the exclusive
production cross
section after each of the CDF cuts.

\begin{table}[hbtp]\begin{center}
\begin{tabular}{|c|c|c|c|c|c|}
\hline
   Mass Fraction Cut & $\nu = $ 0 & $\nu = $ -1 & $\nu = $ -0.5 
                     & $\nu = $ 0.5 & $\nu = $ 1\\
\hline
   $\geq 0.75$ & 14.33 & 194.94 & 52.28 & 3.88 & 0.84 \\
   $\geq 0.8$  & 5.4   & 118.87 & 27.15 & 0.84 & 0.17 \\
   $\geq 0.85$ & 2.02  & 61.89  & 11.13 & 0.17 & 0    \\
   $\geq 0.9$  & 0.34  & 28.43  & 2.87  & 0    & 0    \\
   $\geq 0.95$ & 0.08  & 19.48  & 0.84  & 0    & 0    \\
\hline
\end{tabular}
\caption{\small{Quasi-exclusive cross section (in pb) at the Tevatron,
after CDF cuts,
using different F$_M$ and gluon distributions}.\label{tab:quasitev}}

\begin{tabular}{|c|c|c|c|c|c|}
\hline
   CDF cut & 1 & 2 & 3 & 4 & 5\\
\hline
   Exclusive cross section (pb) & 5.56 $\times 10^3$ & 7.97 $\times 10^2$ & 
5.25 
$\times 10^2$ &
   61.47 & 61.21\\
\hline
\end{tabular}
\caption{\small{Exclusive cross section 
$\sigma_{exc}(p\bar{p} \rightarrow p+\chi_{c0} (\rightarrow J/\psi 
\gamma) + \bar{p})$
(in pb) at the Tevatron energies for
each CDF cut:
1 - one muon with $p_T \geq 1.5$; 
2 - one muon with $p_T \geq 1.5$ and $|\eta| \leq 0.6$;
3 - two muons with $p_T \geq 1.5$;
4 - two muons with $p_T \geq 1.5$ and $|\eta| \leq 0.6$.
5 - same constraint of the forth column plus one gamma with $\eta \leq 3.5$.}
\label{tab:exctev}}

\end{center} \end{table}



\section{\textbf{Possibility of a new measurement at D\O\ }}
\la{VI}
We now examine the possibility of measuring the exclusive $\chic$ production at 
the
Tevatron using the roman pot detectors in the D\O\ collaboration.

After interfacing the DPEMC generator with a program designed to propagate 
(anti)protons
through
the Tevatron lattice, from the D\O\ interacting point to the roman pot 
detectors \cite{kn:fpdacc}, we can predict the number of events observed
for different tagging configurations. 
The Forward Proton Detector (FPD) installed by the D\O\ collaboration
\cite{kn:fpd}  consists of 
eight quadrupole spectrometers, four being located on the
outgoing proton side, and the other four on the antiproton side.

Each spectrometer allows the trajectory reconstruction 
of the outgoing protons and antiprotons near the beam pipe 
determining their energies and scattering angles. 
The quadrupole detectors
are sensitive to outgoing particles with $|t|> 0.6$ GeV$^2$ and 
$\xi < 3. 10^{-2}$, with good acceptance for
high mass objects produced diffractively in the D\O\ main detector. 

We use the following selection cuts:
(($p_T (\mu^{+}) \geq 2.0$ (GeV) or $p_T (\mu^{-}) \geq 2.0$ (GeV)) and
$|\eta (\mu^{\pm})| \leq 2.0$ and $|\eta (\gamma)| \leq 3.0$)
(see Table \ref{tab:acc}).

\begin{table}[hbtp]
\begin{center}
\begin{tabular}{|c|c|c|c|c|}
\hline
  \multicolumn{5}{|c|}{Regular Tevatron Stores - L = 100pb$^{-1}$} \\
\hline
  Scenario & A & B & C & D\\
\hline
   0 & 1.2 $\times 10^8$ & 2.6 $\times 10^6$  &
   4.8 $\times 10^6$ & 2.9 $\times 10^5$\\
   D\O\ selection & 1.8 $\times 10^2$ & 2.7 $\times 10^1$    
   & 3.0 $\times 10^1$ & 1.5 \\
\hline
\end{tabular}
\caption{\small{Number of exclusive $\chi_{c_0}$ events at the Tevatron
(MC error $\sim 10\%$ ) for a regular Tevatron store.
The scenario 0 represents all decay channels included without selection cuts. 
The columns represents the number of events:
A - all (without $p$ or $\bar{p}$ tagging); B - tagged in the $p$ side 
quadrupole; 
C - tagged in the $\bar{p}$ side quadrupole 
and D - double tagged events in the quadrupoles. }
\label{tab:acc}}
\end{center}\end{table}

We note that the number of events in double tagged configuration is quite small 
after applying the selection cuts, so this configuration might not be useful.
However, a single tag event with a rapidity gap on the other side
yields a good number of events, with the additional benefit of having the 
kinematics
determined for one of the scattered particles.


\section{\textbf{Exclusive $\chi_{c_0}$ Production at the LHC}}
\la{VII}
From the results on section IV, we expect a high production cross section of 
$\chi_c$
mesons at the LHC. We can estimate the number of events accessible to the 
TOTEM/CMS detectors, 
benefitting from the good acceptance of the TOTEM detectors for low mass 
objects. 
The TOTEM acceptance for the high $\beta^*$ optics and low $\xi$ values is 
typically 90 \%, 
for the range $0 < |t| < 1$ $GeV^2$.
Then for 10 pb$^{-1}$ of data, 5.3$\times 10^6$ double tagged events
are predicted, with no requirement in the central detector activity. In this 
way, 
one might
look for the $\chi_{c_0}$ in the reconstructed diffractive mass. 

If central activity is required, the lowest possible 
muon $p_T$ cut at low luminosity is on the order of 
$p_T \geq 1.5$ (GeV) for $|\eta| \leq 2.4$ \cite{kn:muoncms}.
Otherwise, the muon $p_T$ threshold
would be $4$ (GeV), which is too high for exclusive $\chi_{c_0}$ production.
The predictions for exclusive and quasi-exclusive production at the LHC are 
shown
in tables \ref{tab:quasilhc} and \ref{tab:exclhc}. We note that the number of
events can be dominated by exclusive production, independent of uncertainties 
in 
the
gluon distribution, if a high
enough cut on the mass fraction can be made, for instance at 0.95, which 
requires 
a good
coverage of the CMS detector at high rapidities.

\begin{table}[hbtp]\begin{center}
\begin{tabular}{|c|c|c|c|c|c|}
\hline
   Mass Fraction Cut & $\nu = $ 0 & $\nu = $ -1 & $\nu = $ -0.5 
                     & $\nu = $ 0.5 & $\nu = $ 1\\
\hline
   $\geq 0.9$  & 1.35 & 138.11  & 17.88  & 0.34 & 0.17  \\
   $\geq 0.95$ & 0    & 13.83   & 1.18   & 0    & 0     \\
\hline
\end{tabular}
\caption{\small{Quasi-exclusive cross section (in pb) at the LHC,
after central activity cuts,
using different mass fractions and gluon distributions, defined in section 
V}.\label{tab:quasilhc}}

\begin{tabular}{|c|c|c|c|c|}
\hline
   Central cut & 1 & 2 & 3 & 4 \\
\hline
   Total & 3.74 $\times 10^3$ & 1.43 $\times 10^3$ & 3.64 $\times 10^2$ &
   1.27 $\times 10^2$ \\
   After Totem Acceptance & 3.03 $\times 10^3$ & 1.16 $\times 10^3$ & 
   2.95 $\times 10^2$ & 1.03 $\times 10^2$ \\
\hline
\end{tabular}
\caption{\small{Exclusive cross section (in pb)
$\sigma_{exc}(p\bar{p} \rightarrow p+\chi_{c_0} (\rightarrow J/\psi 
\gamma) + \bar{p})$
 at the LHC energies for each central cut:
1 - one muon with $p_T \geq 1.5$; 
2 - one muon with $p_T \geq 1.5$ and $|\eta| \leq 2.4$;
3 - two muons with $p_T \geq 1.5$;
4 - two muons with $p_T \geq 1.5$ and $|\eta| \leq 2.4$.
\label{tab:exclhc}}}
\end{center} \end{table}

\section{CONCLUSION}
\la{VIII}

We calculate the diffractive production cross section for $\chi$ mesons at the 
Tevatron and 
LHC using an extended version of the Bialas-Landshoff model, including the full 
kinematics 
needed for low mass states. Both exclusive and inclusive production have been 
evaluated and discussed with various physically motivated cuts at generator 
level. The results can be listed as follows.
\begin{itemize}
\item The results for exclusive production at the Tevatron agree with a recent 
CDF upper limit 
for the exclusive production of $\chi_{c_0},$ with the default parameters
of the model (assuming that the non perturbative gluon coupling
to the proton $G^2/4 \pi \sim 1$);
\item  In the same conditions, the non-exclusive 
background (in particular ``quasi-exclusive'' events which we evaluate from the 
inclusive spectrum) can 
reach similar levels as the exclusive signal, due to experimental and 
theoretical 
uncertainties, 
making difficult to establish the observation of exclusive events within the 
current parameters;
\item  
The possibility of observing exclusive $\chi_{c_0}$ production at the Tevatron, 
using the D\O\ forward 
detector is thoroughly investigated, showing the possibility of a measurement 
if a tight 
cut on the ratio between the $\chi_{c_0}$ mass and the total diffractive mass 
can be performed successfully.
\item Exclusive production  at the LHC, using the CMS/TOTEM detectors, is also 
investigated and appears promising, but again a high
enough cut on the mass fraction is to be made before the observation of 
exclusive production can be established with a good confidence level. It could 
be suitable for a mass fraction, for instance at 0.95, which 
requires 
a good
coverage of the CMS detector at high rapidities.
\end{itemize}

\section{ACKNOWLEDGMENTS}
We would like to thank Maarten Boonekamp for helpful discussions. One of us
(M.S.R) acknowledges support from
CNPq and Capes (Brazil).

\eject
\appendix
\section{\textbf{Exclusive Production kinematics}}
\la{App}
Due the low mass of the $\chi$ mesons, we must evaluate the full kinematics
of an exclusive production. The start point is the
4-momentum conservation in the center-of-momentum frame.
Using the approximation m$_p$ = m$_{\bar{p}}$ = 0, where $p$ and $\bar{p}$ 
represent 
the colliding particles, we can make 
$E_{p,\bar{p}}$ = $|\vec{k}_{p,\bar{p}}|$, and show that:

\begin{equation} \label{eq:conserv}
s=(k_p+k_{\bar{p}})^2-(\vec{k_p}+{\vec{k}_{\bar{p}}})^2+M^2_{diff}+
2E_M(k_p+k_{\bar{p}})-2 \vec{k_M} \cdot (\vec{k_p}+\vec{k}_{\bar{p}})
\end{equation}
where $\vec{k}_{p,\bar{p}}$ is the particle 3-momentum, s
is the center-of-mass energy, $M_{diff}$ is the diffractive mass, and 
$E_M$ and $\vec{k}_M$ are the respective energy and momentum of the produced 
state. 

Due to energy and 3-momentum conservation $E_M = \sqrt{s} - k_p - k_{\bar{p}}$
and $\vec{k_M}=-(\vec{k_p}+\vec{k}_{\bar{p}})$. Moreover, we define

\begin{equation} \label{eq:omega}
\vec{k_p} \cdot \vec{k}_{\bar{p}} \equiv \Omega\ k_p k_{\bar{p}}
\end{equation}

\noindent where $\Omega = -cos{\theta_p} cos{\theta_{\bar{p}}} + sin{\theta_p} 
sin{\theta_{\bar{p}}}(cos{\varphi_p}cos{\varphi_{\bar{p}}}+sin{\varphi_p}
sin{\varphi_{\bar{p}}})$, $\theta$ is the scattering angle and $\varphi$ 
the polar angle.

\noindent Using equation (\ref{eq:omega}) and the conservation constraints in
equation (\ref{eq:conserv}), it can be shown that

\begin{equation} \label{eq:none}
s=2 k_p k_{\bar{p}} (1-\Omega)+M^2_{diff}+2\sqrt{s}(k_p+k_{\bar{p}})-4 
k_p 
k_{\bar{p}}(1-\Omega)
\end{equation}

\noindent Thus
\begin{equation} \label{eq:mass}
M^2_{diff} = s + 2 k_{p} k_{\bar{p}} (1-\Omega) - 2 \sqrt{s} (k_{p} 
+ k_{\bar{p}})
\end{equation}

\noindent using the definition of $\xi$ 

\begin{equation} \label{eq:xidef}
\xi_{p,\bar{p}} = 1 - \frac{k_z^{final}}{k_z^{initial}}\smallskip
\Rightarrow k_{p,\bar{p}} = \frac{\sqrt{s}/2}{cos\theta_{p,\bar{p}}}
(1-\xi_{p,\bar{p}})
\end{equation}

\noindent Using equation (\ref{eq:xidef}) in (\ref{eq:mass}):

\begin{equation} \label{eq:last}
\frac{M^2_{diff}}{s} = 1 +
  \frac{(1-\xi_p)(1-\xi_{\bar{p}})}{2cos\theta_p
  cos_{\bar{p}}}(1-\Omega) - \left ( \frac{1-\xi_p}{cos\theta_p} +
  \frac{1-\xi_{\bar{p}}}{cos\theta_{\bar{p}}} \right )
\end{equation}

\noindent In the case of $|t_{p,\bar{p}}| \to 0$, which means 
$\theta_{p,\bar{p}}\to 0$,
the relation $M^2_{diff} = s \xi_p \xi_{\bar{p}}$ is obtained.

\noindent Note that for convenience in the main text,  the proton kinematics are 
labeled by the index $1$ and the antiproton by $2.$

\eject


\end{document}